\documentclass[epj]{svjour}

\usepackage{lmodern} 
\usepackage[T1]{fontenc}
\usepackage{graphicx,xcolor}
\usepackage{amsmath}
\usepackage{amssymb}
\usepackage{mathrsfs,wasysym}
\usepackage{booktabs}
\usepackage{placeins}
\usepackage{hyperref}

\usepackage{cite}

\newcommand{\mA}{m_A}
\newcommand{\mt}{m_t}

\newcommand{\sss}{\scriptscriptstyle\rm}

\newcommand{\Ord}{\mathcal{O}}
\newcommand{\Lum}{\mathscr{L}}

\newcommand{\as}{\alpha_s}
\newcommand{\muf}{\mu_{\sss F}}
\newcommand{\muh}{\mu_{\sss H}}
\newcommand{\mus}{\mu_{\sss S}}
\newcommand{\mur}{\mu_{\sss R}}

\let\originalleft\left
\let\originalright\right
\renewcommand{\left}{\mathopen{}\mathclose\bgroup\originalleft}
\renewcommand{\right}{\aftergroup\egroup\originalright}
\def\({\left(}
\def\){\right)}
\def\[{\left[}
\def\]{\right]}

\graphicspath{{images/}}

\abstract{
  We consider the production of a pseudo-scalar particle $A$ at the LHC, and present accurate
  theoretical predictions for its inclusive cross section in gluon fusion.
  The prediction is based on combining fixed-order perturbation theory and all-order threshold resummation.
  At fixed order we include the exact next-to-next-to-leading order (NNLO)
  plus an approximate next-to-next-to-next-to-leading order (N$^3$LO$_{\rm A}$) which is
  based on the recent computation at this order for the scalar case.
  We then add threshold resummation at next-to-next-to-next-to leading logarithmic accuracy (N$^3$LL$^\prime$).
  Various forms of threshold resummation are considered, differing by the treatment of subleading terms,
  allowing a robust estimate of the theoretical uncertainties due to missing higher orders.
  With particular attention to pseudo-scalar masses of $200$~GeV and $750$~GeV,
  we also observe that perturbative convergence is much improved when resummation is included.
  Additionally, results obtained with threshold resummation in direct QCD are compared with analogous results
  as computed in soft-collinear effective theory, which turn out to be in good agreement.
  We provide precise predictions for pseudo-scalar inclusive cross section at $13$~TeV LHC for a wide range of masses.
  The results are available through updated versions of the public codes \texttt{ggHiggs} and \texttt{TROLL}.
  \\[1ex]
  OUTP-16-13P
}

\newcommand{\mytitle}{Pseudo-scalar Higgs boson production at N$^3$LO$_{\text{A}}$+N$^3$LL$^\prime$}

\begin{document}

\title{\boldmath\mytitle}

\author{Taushif Ahmed\inst{1}\thanks{\email{taushif@imsc.res.in}} \and Marco Bonvini\inst{2}\thanks{\email{marco.bonvini@physics.ox.ac.uk}} \and M.C. Kumar\inst{3}\thanks{\email{mckumar@iitg.ac.in}} \and Prakash Mathews\inst{4}\thanks{\email{prakash.mathews@saha.ac.in}} \and Narayan Rana\inst{1}\thanks{\email{rana@imsc.res.in}} \and V. Ravindran\inst{1}\thanks{\email{ravindra@imsc.res.in}}\and Luca Rottoli\inst{2}\thanks{\email{luca.rottoli@physics.ox.ac.uk}}}

\institute{The Institute of Mathematical Sciences, IV Cross Road
  CIT Campus, Taramani, Chennai 600113, India \and Rudolf Peierls Centre for Theoretical Physics, 1 Keble Road, University of Oxford, OX1 3NP Oxford, UK \and Department of Physics, Indian Institute of Technology Guwahati, Guwahati 781039, India \and Saha Institute of Nuclear Physics, 1/AF Bidhan Nagar,  Kolkata 700064, India}

\date{\today}


\titlerunning{\mytitle}

\maketitle

\section{Introduction}

The discovery of the Higgs boson by ATLAS \cite{Aad:2012tfa} and CMS
\cite{Chatrchyan:2012ufa} collaborations of the Large Hadron Collider
(LHC) at CERN has put the Standard Model (SM) of particle physics on a
firmer ground. This has led to a better understanding of the dynamics
behind the electroweak symmetry breaking~\cite{Higgs:1964ia,Higgs:1964pj,Higgs:1966ev,Englert:1964et,Guralnik:1964eu}. In
addition, the measured Higgs decay rates~\cite{ATLAS:2013sla,Khachatryan:2014kca} to
$W^+ W^-$, $ZZ$ and heavy fermion pairs are in good agreement with the
predictions of the SM.
Moreover, there are continuous efforts in the 
ongoing $13$ TeV run at the LHC to establish Higgs quantum numbers such
as spin and parity, even though there are already indications that it
is a scalar with even parity~\cite{Aad:2013xqa,Khachatryan:2014kca}.

However, in spite of its spectacular success,
it is well known that the SM fails to explain certain phenomena of the nature such as baryon
asymmetry in the universe, existence of the dark matter, tiny non-zero
mass of the neutrinos, etc.
Explaining these phenomena requires to go beyond the wall of the SM.
Among the several existing models, supersymmetric theories provide an elegant solution to the aforementioned problems.
In one of its simplest realizations, the minimal supersymmetric extension of the SM (MSSM),
the Higgs sector contains two CP-even (scalar), one CP-odd (pseudo-scalar) and two charged Higgs
bosons~\cite{Fayet:1974pd,Fayet:1976et,Fayet:1977yc,Dimopoulos:1981zb,Sakai:1981gr,Inoue:1982pi,Inoue:1983pp,Inoue:1982ej}.
More generally, the existence of additional scalar and pseudo-scalar bosons which couple
to fermions is a prediction of many models which include two Higgs doublets.

If we were to identify the lighter CP-even Higgs boson of these models with the 
observed scalar at the LHC~\cite{Martin:2007pg,Harlander:2008ju,Kant:2010tf},
searches of other Higgs bosons become inevitable.
In particular, for small and moderate $\tan\beta$ (the ratio of vacuum expectation values
$v_1$ and $v_2$ of the two doublets), the large gluon flux at the LHC can provide an
opportunity to search for other Higgs bosons.
There are already efforts along
this direction by the LHC collaborations. However, the experimental
searches crucially depend on precise theoretical predictions.
The goal of this work is to provide precise and accurate theoretical predictions for pseudo-scalar Higgs boson production.

The theoretical predictions for the production of a pseudo-scalar
particle at the LHC have been already available up to NNLO
level~\cite{Harlander:2002vv,Anastasiou:2002wq,Ravindran:2003um} in
perturbative QCD in the heavy top quark limit. These corrections are
large, of the order of 67\% at NLO and an additional 15\%
at NNLO at the central renormalization and factorization scale $\mu_R=\mu_F=\mA/2$, $\mA=200$ GeV.
To achieve sufficient precision, the inclusion of higher orders is therefore necessary.
This situation is very similar to that of scalar Higgs boson production,
for which the N$^3$LO contribution is now known~\cite{Anastasiou:2015ema,Anastasiou:2016cez}.
This is further improved by the resummation of threshold logarithms,
arising from soft gluon emissions, to N$^3$LL$^\prime$ accuracy~\cite{Bonvini:2014joa,Bonvini:2016frm},\footnote
{Here the prime $^\prime$ denotes that in addition to the N$^3$LL terms the resummed result includes
additional (formally higher logarithmic order) terms coming from the matching to N$^3$LO.
For a more detailed discussion, see Refs.~\cite{Bonvini:2014joa,Bonvini:2014tea}. We stress that in most of the literature the prime is omitted.}
leading to a precise determination of the SM Higgs cross section at LHC
with small and reliable uncertainty.

The computation of the
N$^3$LO contribution to pseudo-scalar boson production in the threshold limit has been recently presented in Ref.~\cite{Ahmed:2015qda}.
In this article, we propose a new determination of the pseudo-scalar boson N$^3$LO cross section
based on the recent result at this order for scalar production~\cite{Anastasiou:2016cez}.
Then, we study the inclusion of threshold resummation effects 
to pseudo-scalar production. We do this both in the standard formalism of direct QCD,
as well as in the soft-collinear effective theory (SCET) approach.
We find that the inclusion of threshold resummation together with the approximate N$^3$LO
provides a significant increase of the precision for pseudo-scalar production
and a marked reduction of the theoretical uncertainties.
Our work extends previous results~\cite{deFlorian:2007sr,Schmidt:2015cea} to the next fixed and logarithmic order in QCD.

The structure of this paper is the following.
In Sect.~\ref{sec:pseudoscalar} we introduce the notations and discuss fixed-order results for pseudo-scalar Higgs production.
In Sect.~\ref{sec:resummation} we give an overview of threshold resummation
both in direct QCD and SCET, and present our strategy for the computation of the theoretical uncertainties.
We describe how to construct a precise approximation of the pseudo-scalar Higgs cross section at N$^3$LO in Sect.~\ref{sec:N3LO}.
The numerical impact for pseudo-scalar Higgs production at LHC is presented in Sect.~\ref{sec:results}.
We conclude in Sect.~\ref{sec:conclusions}.

\section{\boldmath Pseudo-scalar production}
\label{sec:pseudoscalar}

The inclusive cross section at hadron colliders with centre of mass
energy $\sqrt{s}$ for the production of a colourless pseudo-scalar particle $A$ of
mass $\mA$ can be written as a convolution
\begin{gather}\label{eq:crosssection}
\sigma(\tau, \mA^2) = \tau \sigma_0\sum_{i,j}\int_{\tau}^1\frac{dz}{z} \Lum_{ij}\(\frac{\tau}{z},\muf^2\) C_{ij}(z,\as,\muf^2)
\end{gather}
of a perturbative partonic coefficient function $C_{ij}(z,\as,\muf^2)$ and a parton luminosity
\begin{gather}
\Lum_{ij}\(x,\muf^2\) = \int_{x}^1\frac{dx'}{x'} f_i\(\frac{x}{x'},\muf^2\) f_j\(x',\muf^2\),
\end{gather}
which is a convolution of parton distribution functions (PDFs) $f_i$ and $f_j$ of the initial state partons $i$ and $j$,
and $\tau =\mA^2/s$.
For simplicity, we assume that $\as=\as(\muf^2)$; computing $\as$ at a different renormalization scale $\mur$
and supplying the coefficients with the corresponding logarithms of the scale is a straightforward task.
The prefactor $\sigma_0$, in case the production is driven by just a top-quark loop with mass $\mt$, reads
\begin{align}
  \label{eq:sigma0}
  \sigma_0 &= \frac{\as^2 G_F}{32 \sqrt{2}\pi} \cot^2\beta \,\big|x_t f(x_t)\big|^2,
  \qquad x_t = \frac{4\mt^2}{\mA^2},\\
  f(x_t) &= 
    \begin{cases}
      {\rm arcsin}^{2}\frac{1}{\sqrt{x_t}} & x_t \geq 1\,, \\
      -\frac{1}{4} \left( \ln
        \frac{1-\sqrt{1-x_t}}{1+\sqrt{1-x_t}} +i \pi
      \right)^{2} & x_t < 1,
    \end{cases}
\end{align}
and is such that $C_{gg}$ is normalized to $\delta(1-z)$ at LO.
In this equation, we assumed a Two Higgs Doublet Model with mixing angle $\beta$.
In the following, we shall not make any assumption on $\beta$, and present results ignoring the $\cot^2\beta$ term:
the resulting cross sections can then be rescaled multiplying by $\cot^2\beta$ to obtain a prediction for any desired value of $\beta$.

The coefficient functions $C_{ij}$ can be computed in perturbation theory.
The NLO~\cite{Kauffman:1993nv,Djouadi:1993ji,Spira:1993bb,Spira:1995rr}
and NNLO~\cite{Harlander:2002vv,Anastasiou:2002wq,Ravindran:2003um}
QCD corrections to the coefficient functions are known
in the large-$\mt$ effective theory, and the NLO also in the exact theory~\cite{Spira:1995rr,Harlander:2005rq}.
Finite $1/\mt$ corrections at NNLO have been computed in Ref.~\cite{Pak:2011hs}.
Threshold contributions at N$^3$LO in the large-$\mt$ limit have been computed in Ref.~\cite{Ahmed:2015qda},
allowing for the computation of an approximate N$^3$LO prediction based on soft-virtual terms.

In this work, we propose a new way of approximating the N$^3$LO contribution,
based on the recent result for scalar Higgs production in the large-$m_t$ effective theory, Ref.~\cite{Anastasiou:2016cez}.
This approximation turns out to be much more precise than any soft-virtual approximation,
and allows us to predict the N$^3$LO cross section for pseudo-scalar Higgs production up to corrections which we expect to be small.
We describe our approximation in Sect.~\ref{sec:N3LO}, after introducing the necessary ingredients in the next Section.

\section{Threshold resummation}
\label{sec:resummation}

We now turn to briefly discussing threshold resummation.
In this work we consider both the standard direct QCD (dQCD) approach~\cite
{Catani:1989ne,Sterman:1986aj,Contopanagos:1996nh,Forte:2002ni,Catani:1996yz}
and the soft-collinear effective theory (SCET) approach~\cite{Ahrens:2008nc,Becher:2006mr,Becher:2007ty,Becher:2006nr}.
We refer the reader to
\cite{Bonvini:2014qga,Sterman:2013nya,Bonvini:2012az,Bonvini:2013td}
for a more detailed discussion about the comparison between the two
frameworks.

Since threshold logarithmic enhancement affects only the gluon-gluon channels, from now on we will focus
on the gluon fusion subprocess, and we will thus drop the parton indices $i$, $j$ assuming they are both equal to $g$. 
Resummation (in dQCD) is usually performed in Mellin space, since the soft-gluon emission phase space factorizes
under Mellin transformation.
The Mellin transformed cross section, Eq.~\eqref{eq:crosssection}, is given by
\begin{gather}
  \sigma(N,\mA^2) \equiv \int_0^1 d\tau \ \tau^{N-2} \sigma(\tau,\mA^2) = \sigma_0 \Lum (N) C(N,\as),
\end{gather}
where we have defined
\begin{align}
  &\Lum(N) \equiv \int_0^1 dz\ z^{N-1} \Lum(z),\\ & C(N,\as) \equiv \int_0^1 dz \ z^{N-1} C(z, \as) 
\end{align}
and for simplicity we have suppressed the dependence on the factorization scale $\muf$.

In $N$ space the threshold limit $z\to 1$ corresponds to the
limit $N\to \infty$. All the non-vanishing contributions to
the coefficient function $C(N,\as)$ can be computed using standard
techniques developed long ago~\cite{Catani:1989ne,Sterman:1986aj,Contopanagos:1996nh,Forte:2002ni,Catani:1996yz},
and one can obtain the all-order resummed coefficient function
\begin{gather}\label{eq:g0expS}
  C_{N\text{-soft}} (N, \as) = g_0 (\as) \exp \mathcal S (\as,\ln N),
\end{gather} 
where $g_0 (\as)$ is a power series in $\as$ and $\mathcal S (\as,\ln N)$ contains purely logarithmically enhanced terms.
This result, which is the standard form of threshold resummation in dQCD,
has been called $N$-soft in Ref.~\cite{Bonvini:2014joa}.
While the function $\mathcal{S}$ needed for N$^3$LL$^\prime$ accuracy has been known for a while~\cite{Moch:2005ba},
as it is identical for pseudo-scalar and scalar Higgs production, the constant function $g_0$ for pseudo-scalar production
was known to second order~\cite{deFlorian:2007sr,Schmidt:2015cea} and it has been computed to third order only recently~\cite{Ahmed:2015qda}.

Besides $N$-soft, there exist several prescriptions, formally
equivalent in the large-$N$ limit, which differ by either power suppressed $1/N$ (subdominant) contributions
or subleading logarithmic terms. We refer the
reader to Ref.~\cite{Bonvini:2014joa} for a more detailed discussion.
In this work, we will use the approach of Ref.~\cite{Bonvini:2016frm},
where it is suggested to vary both subleading and subdominant contributions
to estimate the impact of unknown higher order terms.
Specifically, following Ref.~\cite{Bonvini:2016frm}, we consider the so-called $\psi$-soft
prescription, which essentially amounts to replacing $\ln N\to\psi_0(N)$ in the Sudakov exponent
and performing a collinear improvement.
The resulting default prescription, $\psi$-soft$_2$ (or $\psi$-soft AP2)~\cite{Bonvini:2014joa,Bonvini:2016frm} is given by
\begin{align}\label{eq:psi-soft2}
  C_{\psi\text{-soft}_2} (N, \as) = g_0 (\as) \exp&\Big[ 2\mathcal S (\as,\psi_0(N)) \nonumber \\
                                                 & - 3 \mathcal S (\as,\psi_0(N+1)) \nonumber \\
                                                 & + 2 \mathcal S (\as,\psi_0(N+2))\Big].
\end{align}
The linear combination of shifted exponents implements the collinear improvement AP2,
obtained by retaining the LO splitting function $P_{gg}$ to second order in an expansion in $1-z$.
Alternatively, one can keep only the first order (AP1), leading to
\begin{align}\label{eq:psi-soft1}
  C_{\psi\text{-soft}_1} (N, \as) = g_0 (\as) \exp\mathcal S (\as,\psi_0(N+1)),
\end{align}
which differs from $\psi$-soft$_2$ by subdominant $1/N$ contributions.
Subleading contributions are probed by moving some or all constant terms from $g_0$ to the exponent.
This does not spoil the logarithmic accuracy of the result, but different subleading logarithmic contributions
are generated by interference with the constant terms.
The default position of the constant is determined by retaining in the exponent those constant terms
that naturally arise there from Mellin transformation of threshold logarithms (see Ref.~\cite{Bonvini:2016frm} for further details).
The two variations correspond to either having all constants in the exponent, or no constants in the exponent;
in the latter option all constants are in $g_0$, as in Eqs.~\eqref{eq:psi-soft2}, \eqref{eq:psi-soft1}.

The approach of Ref.~\cite{Bonvini:2016frm} consists then in computing the central value of the resummation
according to $\psi$-soft$_2$ with the default option for the constants,
and the uncertainty on this result from an envelope of scale variation,
variation of $1/N$ terms (AP1 vs AP2) and variation of subleading terms (position of the constants).
This rather conservative procedure for estimating the uncertainty has proved very powerful
in the case of SM Higgs production, where higher order corrections are large and fixed-order scale
uncertainty is an unsatisfactory estimator of missing higher orders~\cite{Bonvini:2016frm}, at least for the first orders.
As we shall see in the next Section, very similar results are found for pseudo-scalar
production, which also suffers from large perturbative corrections.

Alternatively, soft-gluon resummation can be performed in the SCET
framework~\cite{Ahrens:2008nc,Becher:2006mr,Becher:2007ty,Becher:2006nr}.
In this formalism, the partonic coefficient function
$C(z,\as,\muf^2)$ is written in a factorized form as a result of a
sequence of matching steps in which hard and soft modes are subsequently integrated out
\begin{gather}\label{eq:scetfact}
C(z,\as,\muf^2) = H(\muf^2)\, S(z,\muf^2),
\end{gather}
where $H(\muf^2)$ and $S(z,\muf^2)$ are known as hard function and soft function respectively,
and are given as power expansions in $\as$ computed at their last argument.
While the soft function at N$^3$LO is the same for pseudo-scalar and scalar Higgs production
and it has been known for a while~\cite{Li:2014afw,Bonvini:2014tea}, the N$^3$LO hard
function for pseudo-scalar production has been recently computed in Ref.~\cite{Ahmed:2015qpa}.

The hard and soft functions satisfy renormalization group equations
in $\muf$ that can be solved exactly. We can thus write the hard and
soft functions in terms of a hard scale $\muh$ and a soft scale $\mus$, respectively,
by introducing evolution factors which evolve them to the common scale
$\muf$:
\begin{gather}\label{eq:scetres}
C(z,\as,\muf^2) = H(\muh^2)\, S(z,\mus^2)\, U(\muh^2,\mus^2,\muf^2).
\end{gather}
The hard and soft scale should be chosen such that the perturbative expansions of
$H$ and $S$ are well behaved. While for $H$ $\muh\sim \mA$, for the soft function
a typically smaller scale, related to the scale of soft gluon emission, is more appropriate.
Therefore, the evolution $U$ from $\mus$ to $\muf$ performs the resummation of the potentially large logarithms
due to soft radiation.
For the precise choice of scales, we follow the prescription of the original work~\cite{Ahrens:2008nc}.

In this work, we follow Ref.~\cite{Bonvini:2014tea} and consider two independent variations of the SCET resummation:
the variation of subleading $1/N$ terms (corresponding in $z$ space to $(1-z)^0$ terms),
and the inclusion of the so-called $\pi^2$-resummation~\cite
{Ahrens:2008qu,Parisi:1979xd,Magnea:1990zb,Bakulev:2000uh,Eynck:2003fn,Stewart:2013faa}.
As for dQCD, resumming $\pi^2$ constant terms effectively changes subleading terms in the resummation.
On the other hand, the variation of $1/N$ terms is obtained through the inclusion of a collinear improvement,
which effectively amounts to multiplying the soft function by an overall factor $z$~\cite{Bonvini:2014tea}.
This collinear improvement corresponds to the AP1 version of $\psi$-soft.

\section{\boldmath Approximate N$^3$LO cross section}
\label{sec:N3LO}

The recently computed SCET hard function $H$~\cite{Ahmed:2015qpa},
together with the known soft function~\cite{Li:2014afw,Bonvini:2014tea},
allowed the computation of all soft-virtual terms of N$^3$LO pseudo-scalar Higgs production~\cite{Ahmed:2015qda},
i.e.\ the plus distributions terms and the $\delta(1-z)$ term of the coefficient $C_{gg}$.
The quality of such a soft-virtual approximation can be rather good as well as very poor.
The reason is that the soft-virtual terms alone are defined only up to next-to-soft
contributions, i.e.\ terms suppressed by at least one power of $(1-z)$ with respect to the soft ones,
and these next-to-soft terms are usually quite significant~\cite{Ball:2013bra,deFlorian:2014vta,Anastasiou:2014lda}.
Therefore, the quality of any soft-virtual approximation strongly depends on the control one has on the next-to-soft contributions.
Moreover, the soft-virtual approximation only predicts the $gg$ channel,
while other partonic channels, which do not present logarithmic enhancement at threshold, cannot be predicted.
However, other partonic channels, most importantly the $qg$ channel, give a contribution which is non-negligible.
Additionally, including all channels stabilises the factorization scale dependence, which
is instead unbalanced when only the $gg$ channel is included.

In this work we exploit the similarity of pseudo-scalar Higgs production to scalar Higgs production
to provide an approximation which overcomes all the limitations of a soft-virtual approximation.
Calling $C^{H}_{ij}$ the coefficient functions for scalar Higgs production, we can write the
coefficient functions for pseudo-scalar Higgs production as
\begin{equation}\label{eq:deltaCdef}
C_{ij}(z,\as) = \frac{g_0(\as)}{g_0^H(\as)}\Big[C_{ij}^H(z,\as) + \delta C_{ij}(z,\as)\Big],
\end{equation}
where $g_0(\as)$ is the constant function of dQCD resummation for pseudo-scalar Higgs, Eq.~\eqref{eq:g0expS},
and $g_0^H(\as)$ is the analogous function for scalar Higgs.
Eq.~\eqref{eq:deltaCdef} effectively defines $\delta C_{ij}(z,\as)$ as the correction
to the scalar Higgs coefficient functions such that the rescaling $g_0/g_0^H$ converts them
to the pseudo-scalar coefficients.
Expanding order by order in $\as$ both sides of Eq.~\eqref{eq:deltaCdef},
the coefficients $\delta C_{ij}$ at $\Ord(\as^k)$ can be constructed from the knowledge of
the scalar and pseudo-scalar coefficients $C_{ij}$ and $C^H_{ij}$ and of the constant functions $g_0$ and $g^H_0$
up to the same order.
All ingredients are known up to NNLO, allowing the computation of $\delta C_{ij}$ at this order.
At N$^3$LO, $g_0$ and $g^H_0$ are known from resummation~\cite{Ahmed:2015qda,Anastasiou:2014vaa,Bonvini:2014joa,Catani:2014uta}
and $C^H_{ij}$ from Refs.~\cite{Anastasiou:2014lda,Anastasiou:2016cez},
but $C_{ij}$ (and consequently $\delta C_{ij}$) are not known at $\Ord(\as^3)$.
We will argue that using Eq.~\eqref{eq:deltaCdef} to define an approximate
$C_{ij}$ at N$^3$LO by simply setting to zero the unknown $\Ord(\as^3)$ contribution to $\delta C_{ij}$
provides an excellent approximation.

To prove the quality of the approximation, we first observe that if the $\delta C_{ij}$ were unknown
the soft part of the pseudo-scalar coefficients would be predicted exactly by the rescaling
in Eq.~\eqref{eq:deltaCdef}. This observation derives from the fact that in Eq.~\eqref{eq:g0expS}
the Sudakov exponential $\exp \mathcal S$ is identical for scalar and pseudo-scalar production,
and only $g_0$ contains the process dependent part. (This, in turn, also shows that the ratio $g_0/g_0^H$
is identical to the ratio of the SCET hard functions $H$'s for the two processes.)
Therefore, the approximation based on Eq.~\eqref{eq:deltaCdef} is at least as good as a soft-virtual approximation,
as it contains the same information.
In fact, Eq.~\eqref{eq:deltaCdef} contains much more information, thanks to the similarity of the two processes.
To see this, we inspect the $\delta C_{ij}$ terms order by order. Defining the $\as$ expansion as
\begin{equation}
  \label{eq:deltaCas}
  \delta C_{ij}(z,\as) = \frac{\as}{\pi} \delta C_{ij}^{(1)} + \(\frac{\as}{\pi}\)^2 \delta C_{ij}^{(2)}
  + \(\frac{\as}{\pi}\)^3 \delta C_{ij}^{(3)} +\ldots
\end{equation}
we first note that, at NLO,
\begin{equation}
  \label{eq:deltaC1}
  \delta C_{ij}^{(1)} = 0,
\end{equation}
since the difference between scalar and pseudo-scalar production at this order is a pure virtual term~\cite{Spira:1993bb},
and therefore fully accounted for by the rescaling.
Note that this is already highly non-trivial, as by construction $\delta C_{ij}^{(1)}$ has just to be free of soft-virtual contributions;
the fact that the only difference between scalar and pseudo-scalar is corrected by the rescaling
is a clear consequence of the similarity between the two processes considered.
At the NNLO, we find
\begin{align}
  \label{eq:deltaC2}
  \delta C_{gg}^{(2)} &= \frac{495-171z+(20z-2)n_f}{12z}(1-z) \nonumber\\
                      &+\frac{36+21z+2zn_f}{2z}\ln z + \frac{2n_f-27}{3}\ln^2z \nonumber\\
  \delta C_{qg}^{(2)} &= \frac{173-27z}{9z}(1-z) +\frac{24+28z}{3z}\ln z - \frac{28}{9}\ln^2z \nonumber\\
  \delta C_{q\bar q}^{(2)} &= 16\frac{10+12z-(1+z)n_f}{27z}(1-z) \nonumber\\
                      &+32\frac{3+8z-zn_f}{27z}\ln z + \frac{32}{27}\ln^2z \nonumber\\
  \delta C_{qq}^{(2)} &= 8\frac{37-3z}{27z}(1-z) +16\frac{6+11z}{27z}\ln z - \frac{64}{27}\ln^2z \nonumber\\
  \delta C_{qq'}^{(2)} &= 8\frac{11-z}{9z}(1-z) +16\frac{2+3z}{9z}\ln z - \frac{16}{9}\ln^2z.
\end{align}
These results are extremely interesting. We first observe that these terms are next-to-next-to-soft,
namely they are suppressed by $(1-z)^2$ with respect to the leading soft terms (i.e., they vanish in $z=1$).
Moreover, there are no $\ln(1-z)$ terms, which means that those are predicted exactly for any power of $(1-z)$.
Then, we observe that at small-$z$ these expressions are next-to-next-to-leading logarithmic.
Finally, we note that the $\delta C_{ij}$ terms do not contain any explicit scale-dependent contribution at this order.

The fact that the simple rescaling Eq.~\eqref{eq:deltaCdef} allows the prediction of all next-to-soft contributions is very promising:
it shows that the details of the interaction other than those contained in the virtual contributions
are not needed to describe the next-to-soft terms.
This observation, if persisting at higher orders (as we conjecture\footnote
{To support our conjecture, we have tested Eq.~\eqref{eq:deltaCdef} on two
$q\bar q$ dominated processes: Drell-Yan and $b\bar bH$ production.
Also in this case, we find that the $\delta C_{ij}$ coefficients are next-to-next-to-soft,
even though they are non-zero already at NLO, and terms proportional to powers of $\log(1-z)$ appear at NNLO.
}), can be an important step
towards the resummation of next-to-soft contributions~\cite
{Laenen:2008ux,Laenen:2008gt,Laenen:2010uz,Grunberg:2009yi,deFlorian:2014vta,Bonocore:2014wua,Bonocore:2015esa}.
Note that the fact that this is true also for the $qg$ channel is rather informative,
as it tells that the large-$z$ logarithms in this channel, which are formally next-to-soft,
are encoded in the $gg$ subgraphs, as they can be predicted by the knowledge of the virtual $gg$ terms.

\begin{figure}[t]
  \includegraphics[width=0.49\textwidth,page=1]{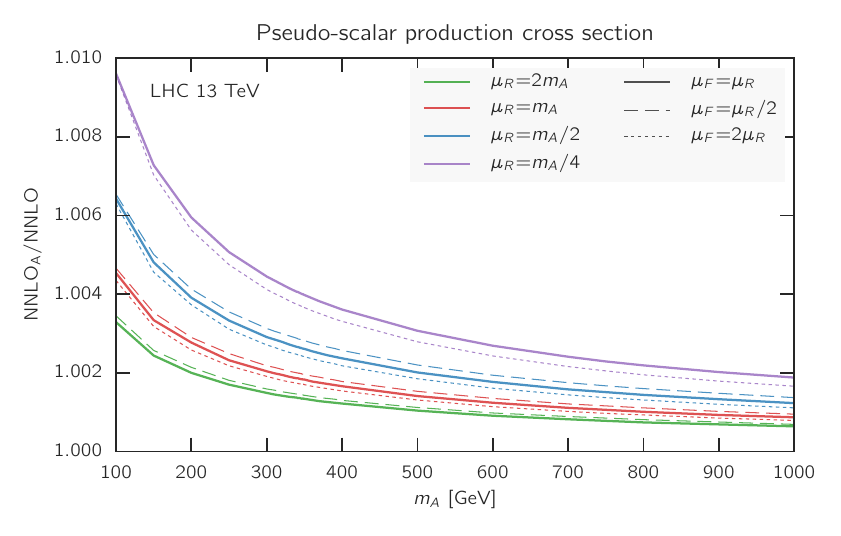}
  \caption{Ratio of approximate NNLO$_{\rm A}$ over exact NNLO pseudo-scalar cross sections,
    as a function of the pseudo-scalar mass $\mA$ at LHC $13$~TeV.
    Curves are shown for four values of $\mur = 2\mA,\mA,\mA/2,\mA/4$ (green, red, blue, purple) and three values
    of $\muf/\mur = 2,1,1/2$ (dotted, solid, dashed).}
  \label{fig:comparisonNNLO}
\end{figure}
 
The other main observation is related to the small-$z$ behaviour.
In the large-$\mt$ effective theory, the leading small-$z$ terms at order $\as^k$ are of the form $(1/z)\ln^{2k-1}z$,
which were shown to coincide between scalar and pseudo-scalar production processes to all orders in $\as$ in Ref.~\cite{Caola:2011wq}.
The absence of next-to-leading logarithmic terms of the form $(1/z)\ln^2z$ in Eq.~\eqref{eq:deltaC2} implies
that, at this order, small-$z$ contributions in the scalar and pseudo-scalar cases start to differ
at the next-to-next-to-leading logarithmic level.
This is perhaps not surprising, as in the effective theory the two largest power of the small-$z$ logarithms, being double logarithms,
are determined by just the hard gluon radiation of the external initial gluon legs.
Hence, we expect this to hold at higher orders as well, thus extending the observation of Ref.~\cite{Caola:2011wq}
to the next-to-leading logarithmic terms.

Therefore, we have found that the rescaling Eq.~\eqref{eq:deltaCdef}, even if the $\delta C_{ij}$ terms are neglected,
reproduces exactly the NLO and deviates from the NNLO by terms which are both next-to-next-to-soft
and next-to-next-to-leading small-$z$, and thus expected to be small.\footnote
{We observe that for scalar Higgs production an approximation based on soft and next-to-soft terms only
is in principle not sufficient to determine the full cross section at high accuracy, neither at N$^3$LO~\cite{Anastasiou:2014lda,Anastasiou:2016cez}
and not even at NLO and NNLO~\cite{Herzog:2014wja}. This derives from the fact that
terms at next-to-next-to-soft and beyond are not uniquely defined, and can be (somewhat artificially)
modified at will by a proper redefinition of the expansion parameters, leading to sizeable effects.
Here, differently, the definition of $\delta C_{ij}$ is unique, and the size of their contribution can only
be assessed by a direct evaluation, which (as we shall see) gives a small effect.
This can be understood by the fact that the rescaling Eq.~\eqref{eq:deltaCdef} also predicts some
(but not all) next-to-next-to-soft terms and higher, and in particular those coming from
the $P_{gg}$ splitting function associated with soft radiation,
which are universal and drive most of the higher soft-order corrections~\cite{Ball:2013bra}.}
To verify this, we plot in Fig.~\ref{fig:comparisonNNLO} the ratio of the approximate NNLO cross section
(denoted NNLO$_{\rm A}$, as obtained setting $\delta C_{ij}=0$) over the exact one, for a range of pseudo-scalar masses and for
various choices of the scales. (The setting of PDFs and other parameters is the same as in Sect.~\ref{sec:results}.)
At high masses, i.e.\ closer to threshold, the difference is at most $\sim2$\textperthousand,
depending on the value of the renormalization scale, but almost independent of the factorization scale
(a consequence of the fact that the factorization scale dependence is generally mild for this process).
At smaller masses, where unpredicted next-to-next-to-soft corrections are larger,
the discrepancy can reach $\sim1\%$ for small renormalization scales.
Overall, the agreement is excellent.

At the next order, we do not have the exact result and therefore we cannot compare.
However, we expect that the $\delta C_{ij}^{(3)}$ coefficients share the same features of the $\delta C_{ij}^{(2)}$,
and as such their contribution should be very small, also considering that the N$^3$LO correction
itself is much smaller than the NNLO one. Numerically, based on the NNLO comparison,
we expect the difference of our approximate N$^3$LO$_{\rm A}$ result to the exact to be just a few permille,
and therefore smaller than scale variation and many other sources of uncertainties.
To further support this expectation, we consider ``variations'' of the approximation itself
to probe the effects of the unknown contributions at N$^3$LO.
The third order coefficient $C_{ij}^{(3)}$ is given, according to Eq.~\eqref{eq:deltaCdef},
and using explicitly Eq.~\eqref{eq:deltaC1}, by
\begin{align}\label{eq:C3}
  C_{ij}^{(3)}(z)
  &= C_{ij}^{H(3)}(z) + r^{(1)} C_{ij}^{H(2)}(z) + r^{(2)} C_{ij}^{H(1)}(z) \nonumber\\
  &+ r^{(3)} C_{ij}^{H(0)}(z) + \delta C_{ij}^{(3)}(z) + r^{(1)}\delta C_{ij}^{(2)}(z),
\end{align}
where $C_{ij}^{H(k)}(z)$ are the expansion coefficients of $C_{ij}^H(z,\as)$ and $r^{(k)}$
are the expansion coefficients of the ratio $g_0(\as)/g_0^H(\as)$.
Our N$^3$LO$_{\rm A}$ is defined by dropping the $\delta C_{ij}^{(3)}(z)$ term in Eq.~\eqref{eq:C3}.
We could equally decide to also drop the last term in the equation,
which would be natural if we had defined $\delta C_{ij}$ differently,
with the rescaling in Eq.~\eqref{eq:deltaCdef} applied only to $C_{ij}^H$ and not to $\delta C_{ij}$.
With this modified definition we obtain a N$^3$LO prediction which only differs by less than $0.3$\textperthousand\ from the N$^3$LO$_{\rm A}$
in the considered range of masses and scales (same as Fig.~\ref{fig:comparisonNNLO}).
This excellent agreement might not be too significant, as it derives from the $\delta C_{ij}^{(2)}(z)$ term,
and can therefore be expected to be roughly the same effect seen at NNLO suppressed by the factor $\as r^{(1)}\sim 0.03$,
so in particular it does not take into account possible larger corrections in the unknown $\delta C_{ij}^{(3)}(z)$ contribution.
\begin{figure}[t]
  \includegraphics[width=0.49\textwidth,page=1]{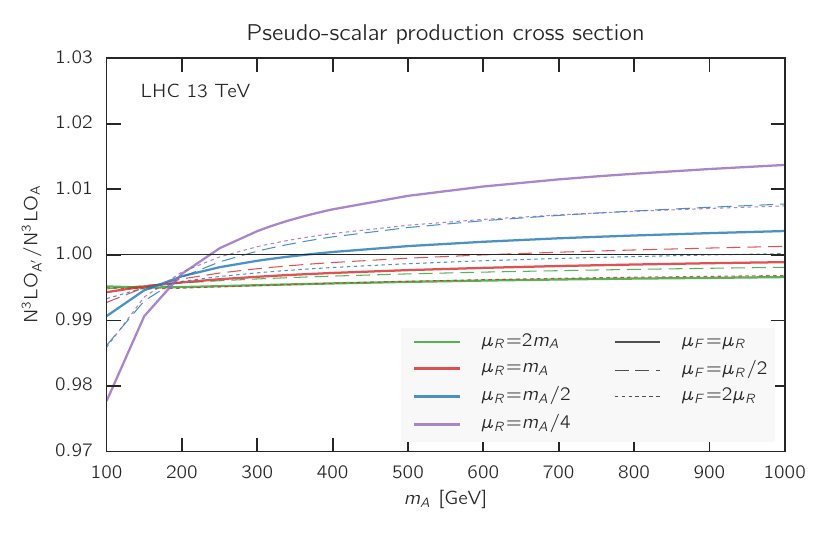}\\
  \includegraphics[width=0.49\textwidth,page=1]{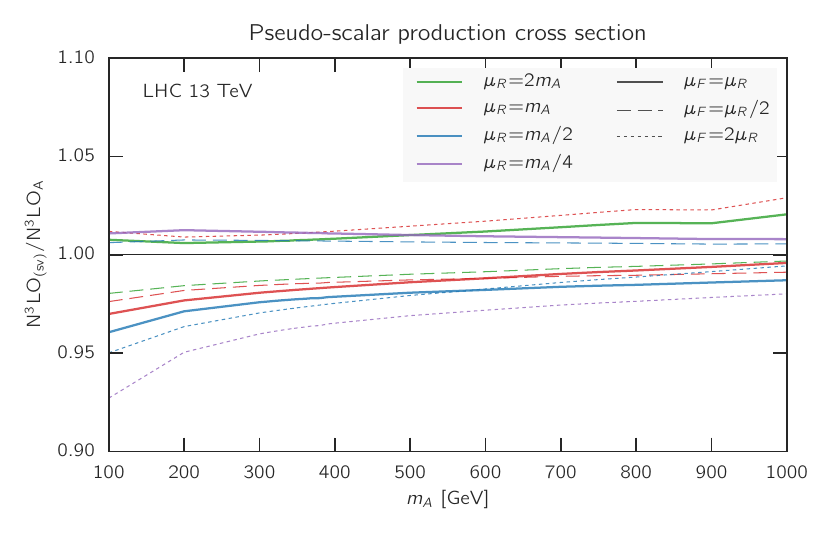}
  \caption{Ratio over approximate N$^3$LO$_{\rm A}$ of the variant approximate N$^3$LO$_{\rm A'}$ described in the text (upper panel)
    and of the soft-virtual N$^3$LO$_{\rm (sv)}$ (lower panel).
    Curves are as in Fig.~\ref{fig:comparisonNNLO}.}
  \label{fig:comparisonN3LO}
\end{figure}
Alternatively, and more drastically, we could ignore the rescaling and drop all the terms in Eq.~\eqref{eq:C3}
except the first: in this case, the ignored terms contain also leading soft and next-to-soft contributions.
Hence, this variation provides a conservative estimate of the error on the approximation,
as it also varies contributions (the soft-virtual ones) which are known and correctly included in our N$^3$LO$_{\rm A}$.
This variation is also useful to understand how big corrections can be if our conjecture on the form of $\delta C_{ij}$,
namely the absence of next-to-soft terms in them, was wrong.
The ratio of this alternative approximation (denoted in the plot N$^3$LO$_{\rm A'}$)
over our default N$^3$LO$_{\rm A}$ is shown in Fig.~\ref{fig:comparisonN3LO} (upper panel),
where it clearly appears that the largest variation never exceeds $2\%$,
and is smaller than $1\%$ for most scales and masses.

Based on these considerations, we would conclude that a realistic
uncertainty on our approximate result is of the order of $1\%$.
In addition, one should also consider the uncertainty coming from the fact that the scalar Higgs N$^3$LO cross section
is itself not known exactly, but as a threshold expansion up to order $(1-z)^{37}$~\cite{Anastasiou:2014lda,Anastasiou:2016cez}.
The uncertainty coming from the truncation of the threshold expansion has been estimated to be
$0.37\%$ for the SM Higgs boson at the $13$~TeV LHC~\cite{Anastasiou:2016cez}.
Since the relative size of the perturbative contributions at various orders is roughly the same for scalar and pseudo-scalar,
this value applies also to our case, for the same mass. At higher masses the process gets closer to threshold,
and the threshold expansion converges more rapidly and is less contaminated by small-$z$ terms
(which are not predicted correctly in the threshold expansion), so the uncertainty from the truncation is likely smaller.
Therefore, the final estimate on the uncertainty on our result remains at the percent level.

\begin{figure}[t]
  \includegraphics[width=0.49\textwidth,page=1]{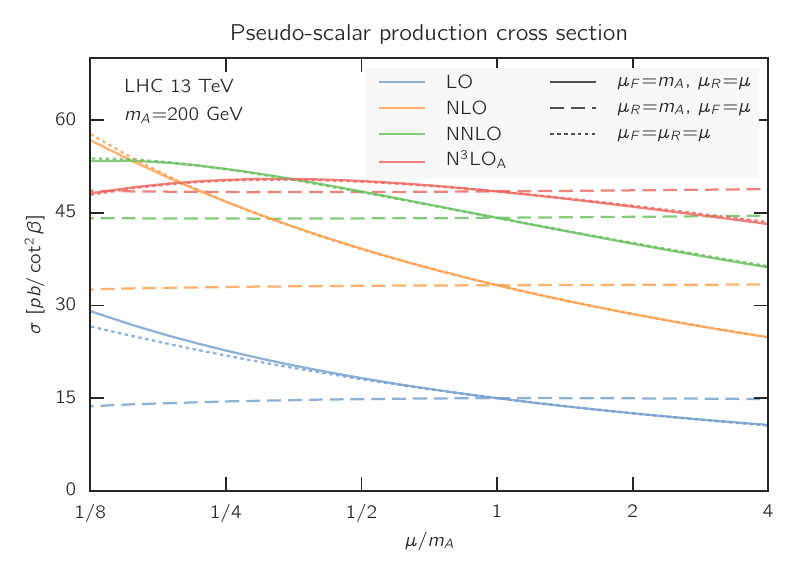}
  \caption{Renormalization (solid), factorization (dashed) and simultaneous (dotted) scale dependence
    for pseudo-scalar production with $\mA=200$~GeV at LHC $13$~TeV. Results at
    LO (blue), NLO (orange), NNLO (green) and N$^3$LO$_{\rm A}$ (red) are shown.}
  \label{fig:scaledep}
\end{figure}

\begin{figure*}[t]
  \includegraphics[width=0.495\textwidth,page=2]{PS_hadr_xsec_res_200_13_uncertainty.pdf}
  \includegraphics[width=0.495\textwidth,page=1]{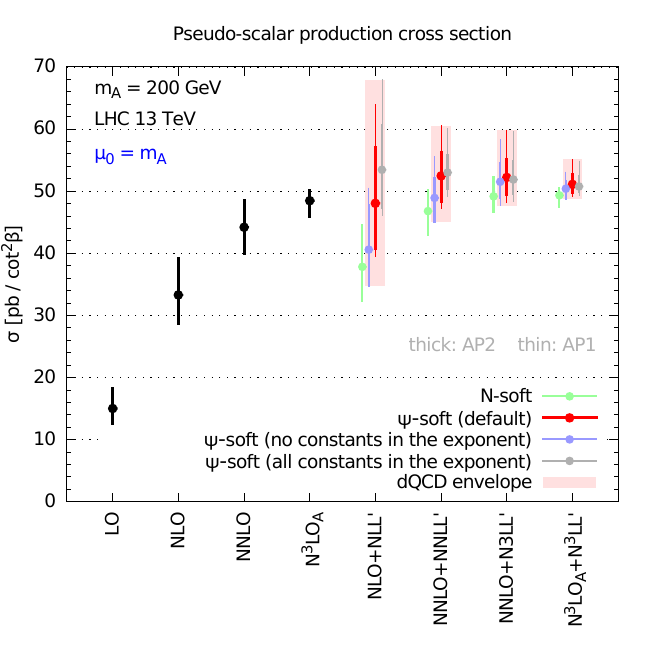}
  \caption{Fixed-order (black) and resummed cross section in dQCD at various orders for $m_A=200$~GeV at the $13$~TeV LHC.
    The standard $N$-soft resummation is shown together with the various variants of $\psi$-soft resummation as discussed in the text.
    The envelope of the various $\psi$-soft resummed results is shown as light-red rectangles.}
  \label{fig:1}
\end{figure*}

We now consider the \emph{modified} soft-virtual (SV) approximation proposed in Ref.~\cite{Ahmed:2015qda},
denoted as N$^3$LO$_{\rm (sv)}$. It consists in approximating the third order coefficient
function by the threshold plus-distributions multiplied by an overall factor $z$.
This approximation proved to be more powerful at previous orders,
and works better in the case of the SM Higgs.
The reason can be traced back to the fact that this modified version includes some collinear contributions,
as proposed in Ref.~\cite{Ball:2013bra}.
Indeed, we notice that this modified SV approximation is close in spirit to the soft$_1$ approximation
of Ref.~\cite{Ball:2013bra}, where $\ln z$ contributions were also retained.
In Fig.~\ref{fig:comparisonN3LO} (lower panel) we plot the ratio of this N$^3$LO$_{\rm (sv)}$
prediction over our N$^3$LO$_{\rm A}$ result, for several choices of scales.
The agreement is typically within 5\%, improving down to 2-3\% at large masses, when the process is closer to threshold
and the soft-virtual approximation is more accurate.
We also not that the $\muf$ dependence is significant, since in the N$^3$LO$_{\rm (sv)}$ it is included only
in the $gg$ channel, and is therefore unbalanced.

To conclude this section, we present the dependence upon renormalization (solid), factorization (dashed) and simultaneous (dotted)
scale variation in Fig.~\ref{fig:scaledep} for LO (blue), NLO (orange), NNLO (green) and N$^3$LO$_{\rm A}$ (red).
We consider a pseudo-scalar mass $\mA=200$~GeV at LHC with $\sqrt{s}=13$~TeV.
While $\muf$ dependence is very flat even at low orders, $\mur$ dependence flattens out significantly at N$^3$LO$_{\rm A}$.
Simultaneous variation of $\mur$ and $\muf$ is very similar to $\mur$ variation.
These results are very similar to those for scalar Higgs production~\cite{Anastasiou:2016cez}.

\section{\boldmath Numerical results at N$^3$LO$_\text{A}$+N$^3$LL$^\prime$}
\label{sec:results}

We now present the results for the inclusive pseudo-scalar cross
section in gluon-gluon fusion at N$^3$LO$_{\rm A}$+N$^3$LL$^\prime$ accuracy at LHC $\sqrt{s}=13$~TeV.
We use the NNLO set of parton distributions \texttt{NNPDF30\_nnlo\_as\_0118}~\cite{Ball:2014uwa}
with $\as = 0.118$ through the LHAPDF 6 interface~\cite{Buckley:2014ana}.
In this study we assume that the pseudo-scalar couples only to top quark and we
take $m_{t}=173.2$~GeV.
We have implemented the exact NNLO and the approximate N$^3$LO$_{\rm A}$ results for pseudo-scalar production
in the public code \texttt{ggHiggs}~\cite{Ball:2013bra,Bonvini:2014jma,Bonvini:2016frm,ggHiggs}.
We then use the public code \texttt{TROLL}~\cite{TROLL,Bonvini:2014joa,Bonvini:2014tea}
to perform the resummation in the dQCD and SCET formalism.

We recall that there have been a series of experimental searches at the LHC for a pseudo-scalar 
boson in gluon fusion as well as bottom associated production channels. 
For instance, the ATLAS collaboration has searched for pseudo-scalar boson over the mass 
window $200\; \text{GeV} < \mA < 1200\; \text{GeV}$ using $13$~TeV data and has put 95\% confidence level (CL) upper limits
on the production cross section times the branching fraction as well as 
95\% CL exclusion limits on the model parameter $\tan\beta$ as a function of $\mA$ 
in different supersymmetric scenarios. For example, with data corresponding to
a luminosity of $3.2$~fb$^{-1}$~\cite{Aaboud:2016cre}, the excluded region is  
$\tan\beta > 7 (47)$ for $\mA = 200 (1000)$~GeV while with luminosity of $13.3$~fb$^{-1}$~\cite{ATLAS:2016fpj} 
the excluded region is $\tan\beta > 9 (42)$ for $\mA = 200 (1200)$~GeV in hMSSM scenarios~\cite{Djouadi:2013uqa}.
Therefore, at the moment, no mass value is excluded, provided the model parameter $\tan\beta$ is in the allowed range.

We first focus on an hypothetical pseudo-scalar mass $\mA=200$~GeV.
In Fig.~\ref{fig:1} we show the inclusive cross section at fixed LO, NLO, NNLO and N$^3$LO$_{\rm A}$ accuracy,
and at NLO+NLL$^\prime$, NNLO+NNLL$^\prime$,
NNLO+N$^3$LL$^\prime$, and N$^3$LO$_{\rm A}$+N$^3$LL$^\prime$ accuracy in
the dQCD approach.
We consider two different values for the central factorization and renormalization scale $\muf=\mur=\mu_0$,
namely $\mu_0=\mA/2$ (left panel) and $\mu_0=\mA$ (right panel).

\begin{figure*}[t]
  \includegraphics[width=0.495\textwidth,page=2]{PS_hadr_xsec_res_200_13_uncertainty_SCET.pdf}
  \includegraphics[width=0.495\textwidth,page=1]{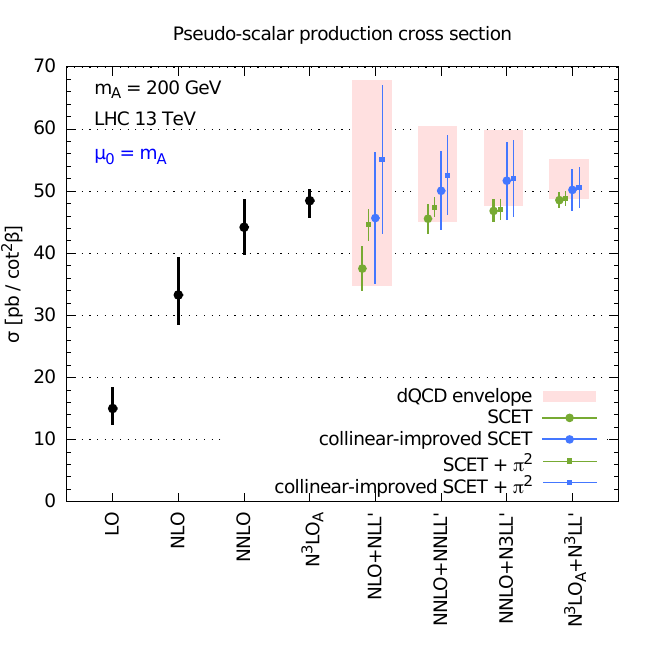}
  \caption{Same as Fig.~\ref{fig:1} but showing the SCET resummed results.
    The dQCD envelope is also shown to facilitate the comparison.}
  \label{fig:2}
\end{figure*}

We included in our results both the NNLO+N$^3$LL$^\prime$ and the N$^3$LO$_{\rm A}$+N$^3$LL$^\prime$ cross sections.
These two constructions have the same fixed order up to $\Ord(\as^2)$, and share the same all-order
resummed contributions from $\Ord(\as^4)$ onwards. However, the contribution
of $\Ord(\as^3)$ is different in the two results: in the N$^3$LO$_{\rm A}$+N$^3$LL$^\prime$
it is given by our approximation of Sect.~\ref{sec:N3LO},
while in the NNLO+N$^3$LL$^\prime$ it is given by the N$^3$LL$^\prime$ resummation expanded to $\Ord(\as^3)$.
In other words, in absence of a full N$^3$LO computation, both provide
alternative ways of estimating the N$^3$LO, which share the same soft, virtual and collinear contributions.
Since in our results we vary the resummation prescription, the NNLO+N$^3$LL$^\prime$
also contains an estimate of the uncertainty on the N$^3$LO itself, and therefore
can be considered as a (much) more conservative estimate of the unknown exact N$^3$LO+N$^3$LL$^\prime$ cross section.

We show the results obtained using different resummation prescriptions. 
Following the approach of Ref~\cite{Bonvini:2016frm}, predictions are 
shown for the $N$-soft and for variants of the $\psi$-soft prescriptions 
which differ by subleading and/or subdominant contributions, as discussed in Sect.~\ref{sec:resummation}.
For each variant we perform a 7-point scale variation varying $\muf$ and 
$\mur$ by a factor 2 up or down and keeping $1/2 \leq \mur/\muf \leq 2$.
The final uncertainty on our predictions is computed as the envelope of the different 
$\psi$-soft variants and each scale variation, and is shown as light-red rectangles in Fig.~\ref{fig:1}.
The uncertainty of the fixed order results is computed as a canonical 7-point
scale variation.

As for the SM Higgs, the fixed order perturbative expansion displays 
poor convergence, especially at lower orders. 
In particular, the NLO correction is more than $100\%$ larger than the LO, and the NNLO 
is a significant correction over the NLO.
Ignoring the LO, which does not contain enough information, we can focus on 
the behaviour of the series at higher orders.
Because of the large perturbative corrections, canonical scale variation does
not guarantee a reliable estimate of the uncertainty from missing higher orders. 
In particular, the NNLO central value is not contained in the NLO uncertainty
band, and the NNLO and the NLO uncertainty bands do not even overlap at 
$\mu_0=\mA$. 
The N$^3$LO$_{\rm A}$ is a smaller correction, perhaps an
indication that the series is finally converging.
The N$^3$LO$_{\rm A}$ value is contained in the NNLO uncertainty 
band, yet is not contained in the NLO uncertainty band; again there is no 
overlap of the two bands.  

Nonetheless, a robust estimate of the missing higher order uncertainty can be 
attained by resorting to resummation. On one hand, resummed results exhibit a
better perturbative behaviour, thereby suggesting that convergence is improved 
when resummed contributions are included. On the other hand, variation of
subleading and subdominant contributions on top of scale variation provides
a more robust method for estimating higher order uncertainty. Contrarily 
to the fixed order, the NLO+NLL$^\prime$ total band fully envelops the 
NNLO+NNLL$^\prime$ band, and the NNLO+N$^3$LL$^\prime$ and
N$^3$LO$_{\rm A}$+N$^3$LL$^\prime$ are contained in the 
NNLO+NNLL$^\prime$ band, which also cover the central value of the
N$^3$LO$_{\rm A}$ result. A similar pattern is observed 
also if only the default $\psi$-soft prescription is considered.
This confirms the conclusions of Ref.~\cite{Bonvini:2016frm} in the context
of SM Higgs production and extends them to the case of pseudo-scalar Higgs production.
Similarly, we also confirm that the central scale choice $\mu_0=\mA/2$ seems a better one,
as it leads to faster convergence and smaller, yet reliable, final uncertainty.

\begin{figure*}[t]
  \includegraphics[width=0.495\textwidth,page=2]{PS_hadr_xsec_res_750_13_uncertainty.pdf}
  \includegraphics[width=0.495\textwidth,page=1]{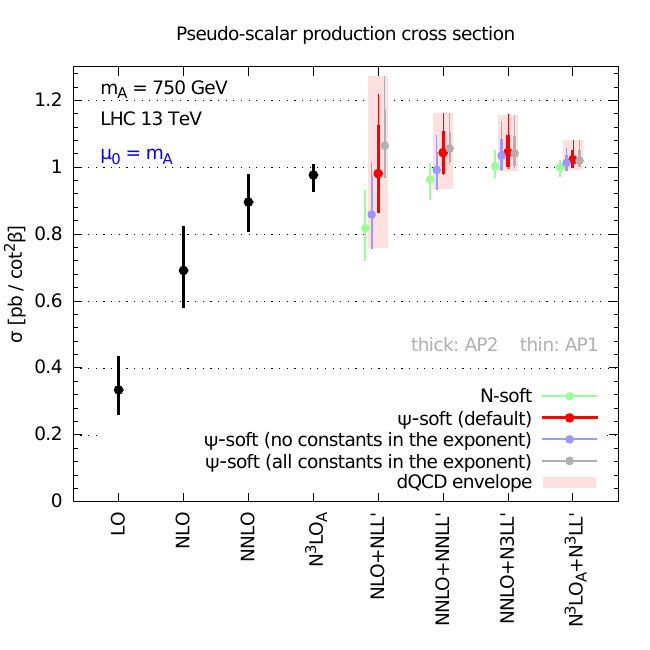}
  \caption{Same as Fig.~\ref{fig:1} but for $m_A=750$~GeV.}
  \label{fig:3}
\end{figure*}

We now analyse the impact of resummation in a framework complementary to
the dQCD approach. In Fig.~\ref{fig:2} we compare the fixed-order results with 
variants of the resummed results obtained in the SCET formalism. 
We perform two different choices of the soft logarithms and we consider the
effect of the $\pi^2$ resummation, as discussed in Sect.~\ref{sec:resummation}.
For each of the variants we compute the uncertainty as in Ref.~\cite{Ahrens:2008nc}.
Specifically, we vary independently $\muf$, $\muh$ and $\mus$, keeping the other
scales fixed when one is varied. As far as $\muf$ and $\muh$ are concerned,
they are varied by a factor of two up and down, about the central scale $\mu_0$,
which we again take to be either $\mu_0=\mA/2$ (left panel) or $\mu_0=\mA$ (right panel).
The definition of the central $\mus$ and of its variation range is more complicated,
 and we refer to Ref.~\cite{Ahrens:2008nc} for a detailed explanation.
For each scale, the largest variation is then symmetrized, and the final
(symmetric) uncertainty is obtained by adding each individual uncertainty in quadrature.
To facilitate the comparison with the dQCD results, in Fig.~\ref{fig:2}
we also show the envelope of the $\psi$-soft variants in dQCD as light-red rectangles. 

We observe that the original formulation of SCET resummation of Ref.~\cite{Ahrens:2008nc} 
leads to a small correction of the fixed order result, due to the choice of the soft logarithms. 
Furthermore, the uncertainty bands are comparable or smaller than their fixed-order counterparts,
suggesting an underestimate of the theory errors.
On the contrary, the impact of resummation is more significant if
subleading terms are included in the form of the collinear improvement of Ref.~\cite{Bonvini:2014tea}.
In this collinear-improved variant the bands are larger and always overlap, indicating a 
better perturbative stability. The inclusion of $\pi^2$ resummation further speeds up
the convergence at $\mu_0=\mA/2$.
The spread of the 
variants we have considered lies almost entirely in the dQCD envelope, with the
exception of the NNLO+N$^3$LL$^\prime$ and N$^3$LO$_{\rm A}$+N$^3$LL$^\prime$ without 
collinear improvement in the $\mu_0=\mA$ case.\footnote
{This is not surprising, since it is known \cite{Ball:2013bra,Bonvini:2014tea} that the choice of
logarithms performed in \cite{Ahrens:2008nc} underestimates the full result if
expanded in powers of $\as$.
Anyway, the difference is not dramatic, and had one symmetrized the dQCD envelope about the central $\psi$-soft
prediction they would be contained in the band.}
Finally, we observe that the central scale $\mu_0=\mA/2$ turns out to be a better choice also from the point of view of SCET
resummation, both because the errors are smaller, and because the impact of higher orders is reduced,
as one can understand from the smaller difference between the original and the collinear-improved versions.

In Fig.~\ref{fig:3} we show the dQCD predictions for a larger pseudo-scalar mass, $m_A=750$~GeV.
This mass value was of some interest in the light of recent measurements~\cite{ATLAS-CONF-2016-018,CMS:2016owr}.
We observe exactly the same pattern found for $m_A=200$~GeV.
The only important difference is that the final uncertainty on the resummed results at $\mu_0=\mA/2$
is smaller than for the lower mass value, probably due to the fact that at larger masses the process is closer
to threshold and the resummation is therefore more accurate (i.e.\ less uncertain) in describing the higher orders.
We do not show the analogous results for SCET resummation, as they have the same features of the lower mass results.
We then conclude that all the observations made for $\mA=200$~GeV remain unchanged for any pseudo-scalar mass.

The comparison of the SCET results with the dQCD ones confirms the procedure suggested in Ref.~\cite{Bonvini:2016frm}
as a robust and reliable method for computing the uncertainty from missing higher orders,
and confirms the scale $\mu_0=\mA/2$ as an optimal central scale.
We can now therefore use this procedure to provide precise and accurate predictions for pseudo-scalar
production at the LHC for generic values of the pseudo-scalar mass $\mA$.

\begin{table}[t]
  \renewcommand{\arraystretch}{1.1}
  \centering
  \begin{tabular}{cc}
    $\mA$[GeV] & $\sigma_{\text{N$^3$LO$_{\rm A}$+N$^3$LL$^\prime$}}$[pb/$\cot^2\beta$] \\
    \midrule
$100$ & $1.71_{-0.08}^{+0.06}\times10^{+2}$ \\
$150$ & $8.29_{-0.32}^{+0.25}\times10^{+1}$ \\
$200$ & $5.03_{-0.16}^{+0.09}\times10^{+1}$ \\
$250$ & $3.64_{-0.10}^{+0.07}\times10^{+1}$ \\
$300$ & $3.22_{-0.09}^{+0.06}\times10^{+1}$ \\
$310$ & $3.27_{-0.08}^{+0.06}\times10^{+1}$ \\
$320$ & $3.39_{-0.09}^{+0.06}\times10^{+1}$ \\
$330$ & $3.66_{-0.09}^{+0.07}\times10^{+1}$ \\
$340$ & $4.28_{-0.11}^{+0.08}\times10^{+1}$ \\
$341$ & $4.39_{-0.11}^{+0.08}\times10^{+1}$ \\
$342$ & $4.53_{-0.11}^{+0.08}\times10^{+1}$ \\
$343$ & $4.69_{-0.12}^{+0.09}\times10^{+1}$ \\
$344$ & $4.90_{-0.12}^{+0.09}\times10^{+1}$ \\
$345$ & $5.18_{-0.13}^{+0.09}\times10^{+1}$ \\
$346$ & $5.68_{-0.14}^{+0.10}\times10^{+1}$ \\
$347$ & $6.33_{-0.16}^{+0.12}\times10^{+1}$ \\
$348$ & $6.24_{-0.15}^{+0.11}\times10^{+1}$ \\
$349$ & $6.16_{-0.15}^{+0.11}\times10^{+1}$ \\
$350$ & $6.07_{-0.15}^{+0.11}\times10^{+1}$ \\
$360$ & $5.28_{-0.13}^{+0.10}\times10^{+1}$ \\
$370$ & $4.60_{-0.11}^{+0.08}\times10^{+1}$ \\
$380$ & $4.02_{-0.10}^{+0.07}\times10^{+1}$ \\
$390$ & $3.53_{-0.08}^{+0.06}\times10^{+1}$ \\
$400$ & $3.10_{-0.07}^{+0.06}\times10^{+1}$ \\
$500$ & $9.71_{-0.20}^{+0.18}\times10^{+0}$ \\
$600$ & $3.60_{-0.07}^{+0.07}\times10^{+0}$ \\
$700$ & $1.51_{-0.03}^{+0.03}\times10^{+0}$ \\
$750$ & $1.01_{-0.02}^{+0.02}\times10^{+0}$ \\
$800$ & $6.89_{-0.11}^{+0.14}\times10^{-1}$ \\
$900$ & $3.38_{-0.05}^{+0.07}\times10^{-1}$ \\
$1000$ & $1.75_{-0.03}^{+0.04}\times10^{-1}$ \\
  \end{tabular}
  \caption{Resummed cross section at N$^3$LO$_{\rm A}$+N$^3$LL$^\prime$ accuracy in dQCD for different values of $\mA$
    at the $13$~TeV LHC.
    The density of $\mA$ values increases close to the $t\bar t$ threshold
    to accurately describe the peak. The error corresponds to the dQCD envelope.}
  \label{tab:mAscan}
\end{table}

\begin{figure}[t]
\centering
  \includegraphics[width=0.495\textwidth,page=1]{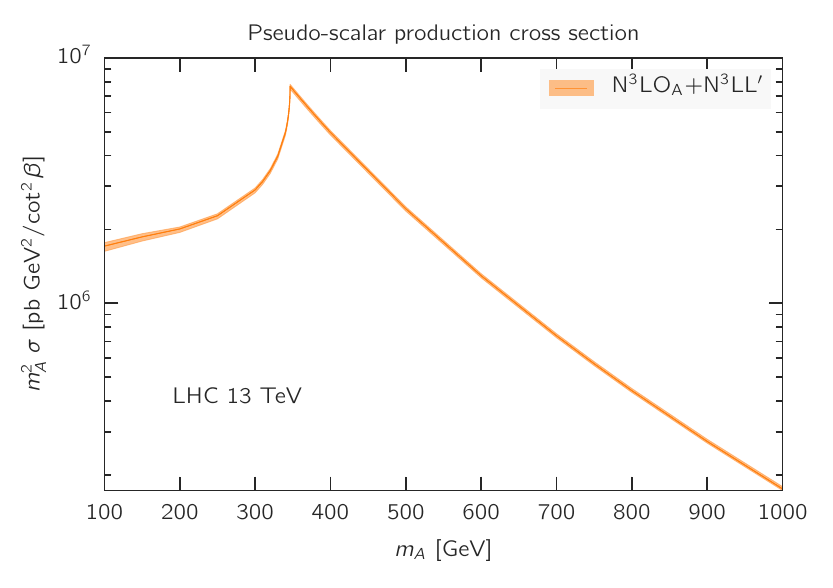}
  \includegraphics[width=0.495\textwidth,page=2]{PS_mA_scan.pdf}
  \caption{Resummed cross section at N$^3$LO$_{\rm A}$+N$^3$LL$^\prime$ accuracy in dQCD as a function of $\mA$ at the $13$~TeV LHC.
    We show both the absolute cross section (upper panel), multiplied by $\mA^2$ for readability,
    and the $K$-factor $\sigma_{\textrm{N$^3$LO$_{\rm A}$+N$^3$LL$^\prime$}}/\sigma_{\textrm{LO}}$ (lower panel).
    In the lower panel we also include the prediction for NNLO+N$^3$LL$^\prime$.
    The error shown corresponds to the dQCD envelope.}
  \label{fig:mAscan}
\end{figure}

In Tab.~\ref{tab:mAscan} we collect the predictions for the inclusive cross section for pseudo-scalar
production at LHC 13 TeV for different values of $\mA$ between $100$~GeV and $1$~TeV.
For each mass value we show predictions at N$^3$LO$_{\rm A}$+N$^3$LL$^\prime$. 
The central value of the resummed result is computed using the default variant of $\psi$-soft$_2$
and the uncertainty is computed as previously discussed,
i.e.\ as the envelope of the different $\psi$-soft variants computed for each scale variation
about the central scale $\mur=\muf=\mA/2$.
The predictions at N$^3$LO$_{\rm A}$+N$^3$LL$^\prime$ are also collected in the form of a plot in Fig.~\ref{fig:mAscan}, 
where we show the resummed cross section and the $K$-factor $\sigma_{\textrm{N$^3$LO$_{\rm A}$+N$^3$LL$^\prime$}}/\sigma_{\textrm{LO}}$
as a function of $m_A$ (orange curve and band).
In the $K$-factor plot we also show, in green, the NNLO+N$^3$LL$^\prime$ uncertainty band.
It is apparent that the knowledge of the N$^3$LO improves significantly the precision of the prediction,
as the error band of the N$^3$LO$_{\rm A}$+N$^3$LL$^\prime$ result is approximately half of the NNLO+N$^3$LL$^\prime$ band.
The latter can be interpreted as a more conservative uncertainty,
covering the uncertainty on the approximate N$^3$LO$_{\rm A}$ result as estimated in Sect.~\ref{sec:N3LO}.

Note that the large-$\mt$ assumption of the effective field theory approach used here is violated for
pseudo-scalar masses $\mA\gtrsim 2\mt$, namely close and after the peak in the upper panel of Fig.~\ref{fig:mAscan}.
However, it is well known (e.g.~\cite{Spira:1995rr,Bonciani:2007ex,Anastasiou:2016hlm}) that the effective theory approach,
when rescaled with the exact LO result, Eq.~\eqref{eq:sigma0},
provides a reasonably good approximation even at large masses,
outside the region of formal validity of the effective approach.
This can be understood in terms of the dominance of soft-collinear contributions, which indeed factorize.
Indeed, the difference between the exact and the effective theory results at NLO reaches
$\sim 10\%$ for $\mA\gtrsim500$~GeV~\cite{Harlander:2012pb,Harlander:2016hcx}, but does not
increase much as $\mA$ gets larger.
The residual effect from missing NNLO finite-$\mt$ terms can then be expected to be a few percent,
as it happens in the scalar case~\cite{Harlander:2009my,Pak:2009dg}.
Therefore, once the known NLO finite-$\mt$ corrections are included,
our results are expected to be reasonably accurate for phenomenology.

\section{Conclusions}
\label{sec:conclusions}

In this work we presented precise predictions for pseudo-scalar Higgs boson production at LHC
based on a combination of fixed order at exact NNLO plus approximate N$^3$LO and of threshold resummation at N$^3$LL$^\prime$.

We have proposed a new method for predicting the N$^3$LO cross section, based on the similarity
of the pseudo-scalar production process in gluon fusion with the analogous scalar production process,
for which the exact N$^3$LO result has been recently made available.
This method consists in a simple rescaling of the perturbative scalar coefficient functions
by the ratio of the process-dependent functions $g_0$ (or hard functions $H$) of the resummation for the two processes.
By construction, this procedure reproduces exactly the soft-virtual-collinear contributions of the pseudo-scalar coefficients.
Interestingly, up to NNLO where the exact result is known, this procedure also reproduces all next-to-soft terms,
all next-to-leading small-$z$ logarithms and all the terms proportional to $\ln(1-z)$ to any positive power.
Assuming this pattern remains true at N$^3$LO and beyond, these observations can also
give some insight on the structure and origin of next-to-soft contributions and how to perform their all-order resummation.
In this work, this allowed us to construct a precise approximation to the N$^3$LO cross section,
up to corrections estimated to be at the percent level.

We then studied the effect of including threshold resummation at N$^3$LL$^\prime$.
We considered threshold resummation both in the traditional direct QCD approach and in the
effective SCET approach. We pay particular attention to the effect of including subleading logarithmic
and subleading power (i.e., beyond threshold) contributions in the resummations.
Following Ref.~\cite{Bonvini:2016frm}, we vary these subleading contributions in dQCD
to obtain a rather conservative uncertainty estimate due to missing higher orders.
This estimate, computed as the envelope of scale and subleading-term variations of the resummed result,
is very reliable, as demonstrated by the fact that the resulting error band successfully covers the next orders.
Specifically, it is much more reliable than the uncertainty estimated by scale variation at fixed order,
which typically underestimates the size of higher order contributions.
Comparison to SCET results further validates the reliability of the dQCD approach.

Differently from the fixed-order results, the resummed results are very stable upon variation
of the central scale, except for the size of the error band which is somewhat dependent on it.
We identify $\mur=\muf=\mA/2$ as an optimal central scale, in the sense that the dQCD error band
turns out to be rather small, but still reliable as demonstrated by the previous orders and the comparison with SCET.
We therefore use this choice to present resummed pseudo-scalar production cross sections
for a wide range of pseudo-scalar masses, from $\mA=100$~GeV to $\mA=1$~TeV.
The $K$-factor with respect to the LO cross section ranges from $\sim3.3$ to $\sim2.3$ respectively,
and the uncertainty estimate from missing higher order ranges from approximately $\pm4\%$ at small mass
to approximately $\pm2\%$ at high mass.
We observe, however, that finite top quark mass effects, neglected in our large-$\mt$ effective theory approach,
become sizeable at large $\mA$. After including the known NLO corrections, the residual effect
from missing NNLO finite $\mt$ contributions can possibly reach a few percent for $\mA\gtrsim 2\mt$.

Our results, although obtained assuming a Two Higgs Doublet Model like the MSSM
for pseudo-scalar boson interactions, can be trivially extended to other more exotic models
by simply changing the Wilson coefficient of the large-$\mt$ effective theory,
which encodes the full-theory information.
The approximate N$^3$LO$_{\rm A}$ is available through the public code \texttt{ggHiggs}~\cite{ggHiggs}, v3.3 onwards,
and the threshold resummation up to N$^3$LL$^\prime$ is available in the public code \texttt{TROLL}~\cite{TROLL}, v3.1 onwards.

\section*{Acknowledgements}
We are grateful to Simone Marzani for interesting discussions and for a critical reading of the manuscript.
We also thank Thomas Gehrmann and Fabio Maltoni for useful discussions.
MB and LR are supported by an European Research Council Starting Grant
``PDF4BSM: Parton Distributions in the Higgs Boson Era''.

\bibliographystyle{JHEP}
\bibliography{pseudo_scalar_res}

\end{document}